# Strain Gradient Elasticity Solution for Functionally Graded Micro-cylinders[1]

H. Sadeghi[a], M. Baghani[b], R. Naghdabadi[b,c]

[a] Department of Mechanical and Aerospace Engineering, University of California, San Diego, La Jolla, CA

[b] Mechanical Engineering Department, Sharif University of Technology, Tehran, Iran 11365-9567

[c] Institute for Nano Science and Technology, Sharif University of Technology, Tehran, Iran

**Abstract**

In this paper, strain gradient elasticity formulation for analysis of FG (Functionally Graded) micro-cylinders is presented. The material properties are assumed to obey a power law in radial direction. The governing differential equation is derived as a fourth order ODE. A power series solution for stresses and displacements in FG micro-cylinders subjected to internal and external pressures is obtained. Numerical examples are presented to study the effect of the characteristic length parameter and FG power index on the displacement field and stress distribution in FG cylinders. It is shown that the characteristic length parameter has a considerable effect on the stress distribution of FG micro-cylinders. Also, increasing material length parameter leads to decrease of the maximum radial and tangential stresses in the cylinder. Furthermore, it is shown that the FG power index has a significant effect on the maximum radial and tangential stresses.

## 1. Introduction

Micro/nano-scale mechanical elements are elements whose characteristic size is in the order of micron or less, e.g., micro/nano-beams, micro/nano-bars and micro/nano-

---

cylinders. These elements are extensively used in micro- and nano-electromechanical systems (MEMS and NEMS) (Huang, Liu et al. 2006; Asghari, Ahmadian et al. 2010; Fu and Zhang 2010; Kahrobaiyan et al. 2011). There are several experimental evidences which show the deformation of such systems is size dependent (Fleck et al. 1994; Stölken and Evans 1998; Lam et al. 2003; McFarland and Colton 2005). Lam, et al. (2003) observed during the bending test of epoxy polymeric beams in micro-scales that the normalized bending rigidity of the beams becomes 2.4 times greater when the thickness of the beam reduces from $115\,\mu m$ to $20\,\mu m$. McFarland and Colton (2005), studied experimentally the bending of polypropylene micro-sized cantilever beams. They observed that the stiffness of the micro-sized cantilevers was at least four times greater than the value which the classical theory of elasticity predicted. Non-classical theories such as couple stress theory, Cosserat continuum, nonlocal elasticity and strain gradient elasticity have been developed recently to consider the size dependency behavior of materials in small-scales (Mindlin and Eshel 1968; Yang et al. 2002; Yoshiyuki 1968). Strain gradient elasticity was originally developed by Mindlin in 1960s (Mindlin 1964; Mindlin 1965; Mindlin and Eshel 1968). In this theory, it is assumed that in addition to strain tensor, gradient of the strain tensor could be taken into account in calculation of the elastic strain energy function. In this way, new characteristic length parameters are introduced and entered into the constitutive equations, in addition to Lame constants. Several studies used the strain gradient elasticity to study the effect of the characteristic length parameters for the materials in micro-scales. Li et al. (2004), used the strain gradient elasticity to obtain a general solution in terms of Fourier integral transforms for a semi-infinite elastic solid with prescribed normal traction. They observed that for constant normal traction, the strain gradient elasticity is capable of predicting the size effects. Paulino et al. (2003), analyzed Mode-III crack problem in Functionally Graded (FG) materials modeled by anisotropic strain gradient elasticity theory. They assumed that the material properties change exponentially and analyzed the effects of gradation parameter as well as the strain gradient parameters on the displacement field near the crack tip. Kong, et al. (2009), studied the static and dynamic problems of Bernoulli-Euler beams analytically on the basis of the strain gradient elasticity theory. They considered two boundary value problems for cantilever beams and analyzed the size effects on the beam bending response and its natural frequencies. Gao and Park (2007) employed a variational formulation and presented an analytical solution for elastic deformation of homogenous thick-walled cylinders using the strain gradient elasticity. Their numerical results revealed that microstructural effects can be large and classical elasticity solution may not be accurate for materials exhibiting significant microstructure dependence. Collin et al. (2009) proposed an analytical solutions for the thick-walled cylinder problem of an isotropic elastic second gradient medium. They showed that in micro-scales the stresses in a homogenous cylinder is completely different from the values predicted by classical elasticity.

Recently, new materials such as nano-materials, piezoelectric materials, shape memory alloys and FG materials, attracted many researchers because of their interesting characteristics. In FG materials, the material properties vary continuously through one or more dimensions. Elastic deformation of FG cylinders has been studied by many researchers (Chen and Lin, 2008; Tutuncu, 2007; Tutuncu and Ozturk, 2001). Tutuncu and Ozturk (2001), obtained a closed form solution for the stresses and displacements in FG cylindrical and spherical vessels subjected to internal pressure. They assumed that material properties vary according to a power law in radial direction. Tutuncu (2007), presented a power series solution for stresses and displacements in FG cylindrical vessels subjected to internal pressure. He assumed an exponentially varying elastic modulus with a constant Poisson ratio and studied the effect of FG power index on the results. Chen and Lin (2008), considered the problem of FG thick-walled cylinders and spherical pressure vessels under pressure loading. They used stress function method to find a general solution for stresses and displacements in thick-walled vessels. They found that the properties of FG materials have a significant influence on stress distribution along the radial direction. Kordkheili and Naghdabadi (2007), presented a semi-analytical thermoelasticity solution for hollow and solid rotating axisymmetric disks. Recently, in addition to the common usages of FGMs, these novel materials are widely used in micro- and nano-structures such as thin films in the form of shape memory alloys (Craciunescu and Wuttig 2003; Fu, Du et al. 2003), micro- and also nano-electromechanical systems (MEMS and NEMS) (Fu, Du et al. 2004; Witvrouw and Mehta 2005). Since most of the studies on FG cylinders have used the classical theory of elasticity, it seems necessary to have an analytical solution for elastic analysis of FG micro-cylinders because of their potential application in the micro-electromechanical systems.

The purpose of this study is to consider the effect of characteristic length parameter on the elastic deformation of axisymmetric FG micro-cylinders. In this way, the strain gradient elasticity formulation for FG micro-cylinders is presented. The displacement form of the governing equation is given and power series solutions for stresses and displacements are obtained. Numerical examples are solved for different values of the characteristic length parameter and the results are compared with those of the classical elasticity. Furthermore, the effect of FG power index on the stresses is studied.

## 2. Strain Gradient Elasticity Formulation

### 2.1. Governing Equation

In the strain gradient elasticity the elastic strain energy density function is written as a function of the gradient of strain tensor, in addition to the strain tensor. The simplified form of the elastic strain energy density function presented by Altan and Aifantis (1997), is used in this study. Thus, the elastic strain energy density function for a FG material can be written as

$$w = w(\varepsilon_{ij}, \varepsilon_{ij,k}) = \frac{1}{2}\lambda(x,y)\varepsilon_{ii}\varepsilon_{jj} + \mu(x,y)\varepsilon_{ij}\varepsilon_{ij} + l^2\left(\frac{1}{2}\lambda(x,y)\varepsilon_{ii,k}\varepsilon_{jj,k} + \mu(x,y)\varepsilon_{ij,k}\varepsilon_{ji,k}\right) \quad (1)$$

where $w$ is the elastic strain energy density function, $\lambda(x,y)$ and $\mu(x,y)$ are Lame parameters, $l$ is the characteristic length parameter, $\varepsilon_{ij} = \frac{1}{2}(u_{i,j} + u_{j,i})$ are components of strain tensor, in which $u_i$ are components of displacement vector and $u_{i,j}$ stand for the derivatives of $u_i$ with respect to $x_j$. In equation (1) Lame parameters are assumed to be function of two coordinates $x$ and $y$. For an axisymmetric FG cylinder, elastic strain energy density function can be written in the cylindrical coordinates as

$$w = \frac{1}{2}\lambda(r)\left(\varepsilon_{rr}(r)+\varepsilon_{\theta\theta}(r)+\varepsilon_{zz}(r)\right)^2 + \mu(r)\left(\varepsilon_{rr}^2(r)+\varepsilon_{\theta\theta}^2(r)+\varepsilon_{zz}^2(r)\right) + \frac{1}{2}\lambda(r)l^2\left(\frac{d}{dr}\varepsilon_{rr}(r)+\frac{d}{dr}\varepsilon_{\theta\theta}(r)+\frac{d}{dr}\varepsilon_{zz}(r)\right)^2$$
$$+ \mu(r)l^2\left[\left(\frac{d}{dr}\varepsilon_{rr}(r)\right)^2 + \left(\frac{d}{dr}\varepsilon_{\theta\theta}(r)\right)^2 + \left(\frac{d}{dr}\varepsilon_{zz}(r)\right)^2\right] \quad (2)$$

where Lame parameters are assumed to be functions of radial position, $r$. The components of Cauchy stress tensor, $\tau_{ij}$, the double stress tensor, $\mu_{ijk}$, the strain gradient tensor, $\kappa_{ijk}$, and total stress tensor, $\sigma_{ij}$, are respectively given by Altan and Aifantis (1997).

$$\tau_{ij} = \frac{\partial w}{\partial \varepsilon_{ij}} = \tau_{ji} \quad (3)$$

$$\mu_{ijk} = \frac{\partial w}{\partial \kappa_{ijk}} = \mu_{jik} \quad (4)$$

$$\kappa_{ijk} = \varepsilon_{ij,k} = \frac{1}{2}(u_{i,jk} + u_{j,ik}) \quad (5)$$

$$\sigma_{ij} = \tau_{ij} - \mu_{ijk,k} \quad (6)$$

Also, the equilibrium equation in the simplified strain gradient elasticity can be given by Altan and Aifantis (1997).

$$\sigma_{ij,j} + f_i = 0 \quad (7)$$

In the cylindrical coordinates, radial and tangential strains, $\varepsilon_{rr}$ and $\varepsilon_{\theta\theta}$, can be written in terms of radial displacement $u(r)$ as

$$\varepsilon_r = \frac{du(r)}{dr}, \qquad \varepsilon_\theta = \frac{u(r)}{r} \quad (8)$$

In this paper it is assumed that Lame constants vary along the thickness according to the power law

$$\lambda(r) = \lambda_0 r^n, \qquad \mu(r) = \mu_0 r^n \quad (9)$$

where $\lambda_0$ and $\mu_0$ are material constants and $n$ is FG power index. The equilibrium equation in the cylindrical coordinate system for axisymmetric condition is

$$\frac{d\sigma_{rr}}{dr} + \frac{\sigma_{rr} - \sigma_{\theta\theta}}{r} = 0 \quad (10)$$

Combining equations (2)-(4) with equations (8)-(10) the governing equation for an axisymmetric long FG cylinder in the strain gradient elasticity is obtained in terms of radial displacement $u(r)$ as follows

$$[f_4(r)\frac{d^4}{dr^4}+f_3(r)\frac{d^3}{dr^3}+f_2(r)\frac{d^2}{dr^2}+f_1(r)\frac{d}{dr}+f_0(r)]\,u(r)=0 \tag{11}$$

in which $f_0(r)$ to $f_5(r)$ are known functions defined as

$$\begin{aligned}
f_0(r) &= \left(\mu_0(4-2n)+\lambda_0(n-2)(n-3)\right)l^2 r^{-4}+\left(\lambda_0(n-1)-2\mu_0\right)r^{-2}\\
f_1(r) &= (\lambda_0+2\mu_0)(n+1)r^{-1}+\left(-\lambda_0(n-2)(n-3)+\mu_0(2n-4)\right)l^2 r^{-3}\\
f_2(r) &= (\lambda_0+2\mu_0)+\left(2\mu_0(1-n^2)-\lambda_0(n^2+n-3)\right)l^2 r^{-2}\\
f_3(r) &= -(\lambda_0+2\mu_0)(1+2n)l^2 r^{-1}\\
f_4(r) &= -(\lambda_0+2\mu_0)l^2
\end{aligned} \tag{12}$$

It is noted that the governing equation for an axisymmetric long FG cylinder in strain gradient elasticity is a fourth order ODE whereas in the classical elasticity such equation is a second order ODE. Also, it is understood that in the absence of strain gradient effect ($l=0$) the governing equation in strain gradient elasticity reduces to the same second order governing equation in the classical elasticity.

## 2.2 Boundary Conditions

It is assumed that the FG cylinder is subjected to boundary pressures $P_i$ and $P_o$ on its inner and outer surfaces, respectively. So, the boundary conditions for the proposed problem can be expressed as Altan and Aifantis (1997).

$$\begin{aligned}
\left\{\tau_{rr}-l^2\left[\frac{d^2\tau_{rr}}{dr^2}+\frac{1}{r}\left(\frac{d\tau_{rr}}{dr}-\frac{d\tau_{\theta\theta}}{dr}\right)-\frac{2}{r^2}(\tau_{rr}-\tau_{\theta\theta})\right]\right\}_{r=r_i} &= -P_i\\
\left\{\tau_{rr}-l^2\left[\frac{d^2\tau_{rr}}{dr^2}+\frac{1}{r}\left(\frac{d\tau_{rr}}{dr}-\frac{d\tau_{\theta\theta}}{dr}\right)-\frac{2}{r^2}(\tau_{rr}-\tau_{\theta\theta})\right]\right\}_{r=r_o} &= -P_o\\
\left. l^2\frac{d\tau_{rr}}{dr}\right|_{r=r_i} &= 0\\
\left. l^2\frac{d\tau_{rr}}{dr}\right|_{r=r_o} &= 0
\end{aligned} \tag{13}$$

The prescribed double stress traction is considered zero on both the inner and outer surfaces. It can be seen that in the absence of the strain gradient effect ($l=0$) these four boundary conditions reduce to the two boundary conditions in the classical theory of elasticity.

## 3. Solution Technique

In this section a series solution will be presented for the governing equation (11). Because $r=0$ is a regular singular point of the governing equation (11), the solution of this equation can be expressed in the series form as

$$u(r) = \sum_{j=0}^{\infty} w_j r^{t+j} \tag{14}$$

Substituting equation (14) in the governing equation (11) and by doing some simplifications, the following recurrence formula is obtained

$$w_j = -\frac{A_j}{B_j} w_{j-2} \; ; \qquad \text{for } j=2,4,6,\ldots$$

$$w_j = 0 \; ; \qquad \text{for } j=1,3,5,\ldots \tag{15}$$

where

$$\begin{aligned}
A_j &= (\lambda_0 + 2\mu_0)(n+1)(t+j-2) + (\lambda_0 + 2\mu_0)(t+j-2)(t+j-3) + (n\lambda_0 - 2\mu_0 - \lambda_0) \\
B_j &= \left(4\mu_0 + n^2\lambda_0 - 2n\mu_0 - 5n\lambda_0 + 6\lambda_0\right)l^2 + \left(-6\lambda_0 + 2n\mu_0 + 5n\lambda_0 - \lambda_0 n^2 - 4\mu_0\right)(t+j)l^2 \\
&\quad + \left(2\mu_0 - n^2(\lambda_0 + 2\mu_0) + 3\lambda_0 - n\lambda_0\right)(t+j)(t+j-1)l^2 \\
&\quad - (\lambda_0 + 2\mu_0)(1+2n)(t+j)(t+j-1)(t+j-2)l^2 \\
&\quad - (\lambda_0 + 2\mu_0)(t+j)(t+j-1)(t+j-2)(t+j-3)l^2
\end{aligned} \tag{16}$$

Substituting equation (14) into equation (11) and doing some simplifications, we obtain the characteristic equation of the FG cylinder. For the sake of brevity, the final form is presented as follows.

$$\begin{aligned}
&(\lambda_0 + 2\mu_0)t^4 + (\lambda_0 + 2\mu_0)(5-2n)t^3 + \left(-n^2(\lambda_0 + 2\mu_0) + 5\lambda_0(n-1) + \mu_0(12n-14)\right)t^2 \\
&+ \left(\lambda_0(2n-5) + \mu_0(2+2n^2-6n)\right)t + \lambda_0(n^2 - 5n + 6) + \mu_0(4-2n) = 0
\end{aligned} \tag{17}$$

This is a fourth order algebraic equation in terms of $t$ and has four roots

$$\begin{aligned}
t_1 &= 1, \\
t_2 &= 2-n, \\
t_3 &= -\frac{1}{2}\left(n - 2 - \frac{\sqrt{(\lambda_0 + 2\mu_0)^2 n^2 + (8\lambda_0^2 + 24\mu_0\lambda_0 + 16\mu_0^2)(2-n)}}{\lambda_0 + 2\mu_0}\right), \\
t_4 &= -\frac{1}{2}\left(n - 2 + \frac{\sqrt{(\lambda_0 + 2\mu_0)^2 n^2 + (8\lambda_0^2 + 24\mu_0\lambda_0 + 16\mu_0^2)(2-n)}}{\lambda_0 + 2\mu_0}\right)
\end{aligned} \tag{18}$$

Depending on the relation between $t_r$'s the solution of the governing equation (11) can be found. Here, without losing the generality of the solution, we just consider the more general case in which $t_p - t_q$ ($p, q = 1, ..., 4$, $p \neq q$) are not integer numbers. In this case the general solution of the governing equation (11) can be written as

$$u(r) = \sum_{i=1}^{4} c_i u_i = \sum_{i=1}^{4} c_i \sum_{j=1}^{\infty} w_j r^{t_i + j} \qquad (19)$$

where $c_i$ ($i=1,...,4$) are unknown constants to be determined from the boundary conditions. For the special case in which $t_p - t_q$ ($p, q = 1, ..., 4$, $p \neq q$) are integer numbers, $u_p$ and $u_q$ can be obtained as

$$u_p(r) = \sum_{m=1}^{\infty} w_m r^{m+t_p}$$

$$u_q(r) = u_p(r) \ln(r) + \sum_{m=1}^{\infty} w_m r^{m+t_q} \qquad (20)$$

Since, the cases in which $t_p - t_q$ ($p, q = 1, ..., 4$, $p \neq q$) are not integer numbers, are more general than the other cases, we will continue the calculations for this case and for the other cases results can be found in the same way. So, Cauchy stresses can be found as

$$\tau_{rr} = \sum_{i=1}^{4} \left( \lambda_0 w_0 c_i r^{t_i + n - 1} + \sum_{j=1}^{\infty} c_i w_j \left[ (\lambda_0 + 2\mu_0)(t_i + j) + \lambda_0 \right] r^{t_i + j + n - 1} \right)$$

$$\tau_{\theta\theta} = \sum_{i=1}^{4} \left( (\lambda_0 + 2\mu_0) w_0 c_i r^{t_i + n - 1} + \sum_{j=1}^{\infty} c_i w_j \left[ (\lambda_0 + 2\mu_0) + \lambda_0 (t_i + j) \right] r^{t_i + j + n - 1} \right) \qquad (21)$$

To find the unknown constants $c_i$ ($i=1..4$), Cauchy stresses should be substituted from equations (21) into the boundary conditions (13) that result in a set of algebraic equations which can be written in matrix form as

$$[K] \{ c_1 \; c_2 \; c_3 \; c_4 \}^T = [F] \qquad (22)$$

where and $[F]$ is

$$[F] = \{ -P_i \; -P_o \; 0 \; 0 \}^T \qquad (23)$$

and for $i = 1, ..., 4$, $[K]$ matrix can be given as

$$K_{1,i} = \sum_{j=1}^{\infty} \left( \begin{array}{c} (\lambda_0 + 2\mu_0) \left( \begin{array}{c} 2j^2 - j^3 - t_i^3 - 2t_i^2 n - 3jt_i^2 - t_i n^2 - 3t_i j^2 \\ -jn^2 + 2t_i^2 + 4t_i j - 4t_i nj + t_i + j + jn + t_i n \end{array} \right) \\ + \lambda_0 n(3-n) + \mu_0 (2t_i + 2jn + 2j - 2 + 2t_i n - 4nj^2 + 2n) \end{array} \right) l^2 w_j r_1^{t_i + n + j - 3}$$

$$+ \sum_{j=1}^{\infty} w_j \left( (\lambda_0 + 2\mu_0)(t_i + j)\lambda_0 \right) r_1^{t_i + n + j - 1} + w_0 \left( \lambda_0 r_1^{t_i + n - 1} + l^2 \left( \begin{array}{c} \lambda_0 \left( \begin{array}{c} -t_i^2 + (3 - 2n)t_i \\ -n^2 + 3n - 2 \end{array} \right) \\ + \mu_0 (2n + 2t_i - 6) \end{array} \right) r_1^{t_i + n - 3} \right)$$

$$K_{2,i} = \sum_{j=1}^{\infty} \left( \begin{array}{c} (\lambda_0 + 2\mu_0) \left( \begin{array}{c} 2j^2 - j^3 - t_i^3 - 2t_i^2 n - 3jt_i^2 - t_i n^2 \\ -3t_i j^2 - jn^2 + 2t_i^2 + 4t_i j - 4t_i nj + t_i + j + jn + t_i n \end{array} \right) \\ + \lambda_0 n(3-n) + \mu_0 (2t_i + 2jn + 2j - 2 + 2t_i n - 4nj^2 + 2n) \end{array} \right) l^2 w_j r_2^{t_i + n + j - 3} \quad (24)$$

$$+ \sum_{j=1}^{\infty} w_j \left( (\lambda_0 + 2\mu_0)(t_i + j)\lambda_0 \right) r_2^{t_i + n + j - 1} + w_0 \left( \lambda_0 r_2^{t_i + n - 1} + l^2 \left( \begin{array}{c} \lambda_0 \left( \begin{array}{c} -t_i^2 + (3 - 2n)t_i \\ -n^2 + 3n - 2 \end{array} \right) \\ + \mu_0 (2n + 2t_i - 6) \end{array} \right) r_2^{t_i + n - 3} \right)$$

$$K_{3,i} = \lambda_0 w_0 (t_i + n - 1) r_1^{t_i + n + j - 2} + \sum_{j=1}^{\infty} w_j \left( (\lambda_0 + 2\mu_0)(t_i + j) + \lambda_0 \right)(t_i + n + j - 1) r_1^{t_i + n + j - 2}$$

$$K_{4,i} = \lambda_0 w_0 (t_i + n - 1) r_2^{t_i + n + j - 2} + \sum_{j=1}^{\infty} w_j \left( (\lambda_0 + 2\mu_0)(t_i + j) + \lambda_0 \right)(t_i + n + j - 1) r_2^{t_i + n + j - 2}$$

So, the unknown coefficients $c_i$ ($i = 1,...,4$) can be found simply as

$$\{c_1 \quad c_2 \quad c_3 \quad c_4\}^T = [K]^{-1}[F] \quad (25)$$

## 4. Numerical Results and Discussion

In this section, numerical examples will be presented to study the effect of characteristic length parameter and FG power index on the elastic response of FG micro-cylinders. For verification of the proposed solution, a homogenous micro-cylinder is considered with the inner and outer radii $r_i = 1\mu m$ and $r_o = 5\mu m$, respectively. It is assumed that the cylinder is subjected to boundary pressures $P_i = 10 MPa$ and $P_o = 0$ at the inner and outer surfaces. Fig. 1 shows a plot of the non-dimensional radial stress for the homogenous micro-cylinder along the radius and the result is compared with the results given by Gao and Park (2007) in which the following parameter values are used in the calculations $l = 0.5 \mu m$, $E = 135 GPa$ and $v = 0.3$. Also, the results obtained from the classical theory of elasticity are presented in this figure. It can be seen that the results are in good agreement with those obtained from the solution presented in Gao and Park (2007) for a homogenous micro-cylinder.

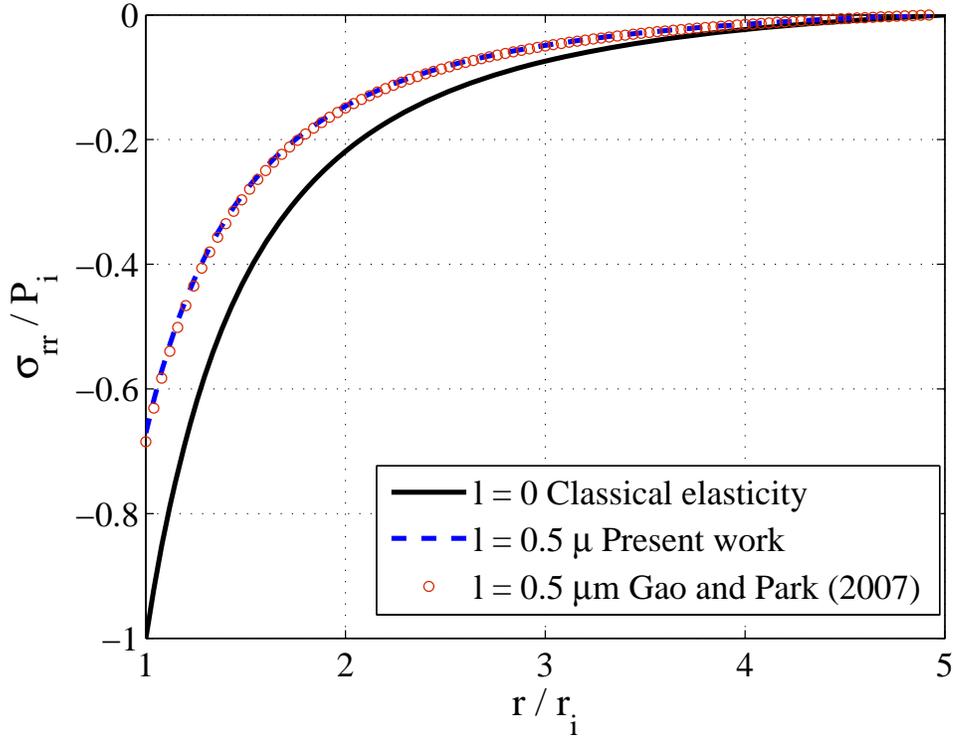

Figure 1: Comparison of non-dimensional radial stress along the radius of homogenous micro-cylinder with $r_i = 1\mu m$, $r_o = 5\mu m$, $l = 0.5\mu m$, $E = 135 \text{GPa}$ and $\nu = 0.3$

Consider a FG micro-cylinder with inner and outer radii $r_i = 1\ \mu m$ and $r_o = 5\ \mu m$, respectively. It is assumed that the material properties are $\lambda_i = 20$ GPa and $\mu_i = 10$ GPa at the inner radius of the cylinder and the cylinder is subjected to internal pressure $P_i = 10 \text{MPa}$. The distribution of non-dimensional radial displacement as well as non-dimensional radial and hoop stresses along the cylinder radius in a micro-cylinder with FG power index $n = 0.5$ is given in Fig. 2-Fig. 4 for some values of the characteristic length parameter, *l*. In these figures, the classical elasticity solution ($l = 0$) is also shown for comparison. It can be seen that the material characteristic length parameter has a significant effect on the stress distribution in the FG micro-cylinder and as the material length parameter decreases radial and tangential stresses coincide with the results obtained from the classical elasticity. Also, it can be understood that increasing the material length parameter decreases the maximum radial and tangential stresses. For instance, increasing material length parameter from $l=0$ to $l=0.4$ $\mu m$ decreases the maximum radial and tangential stresses %52 and %78, respectively. Furthermore, it can be observed that the variance between the values calculated by strain gradient elasticity and classical elasticity are greater at the inner radius of the cylinder, rather than the outer radius. For example, there is a difference of %90 between the values of non-dimensional tangential stresses predicted by strain gradient elasticity with $l=0.4$ $\mu m$ and those obtained by the classical elasticity at the inner radius of the cylinder. Whereas, there is %33 difference between the values calculated by these two theories at the outer radius.

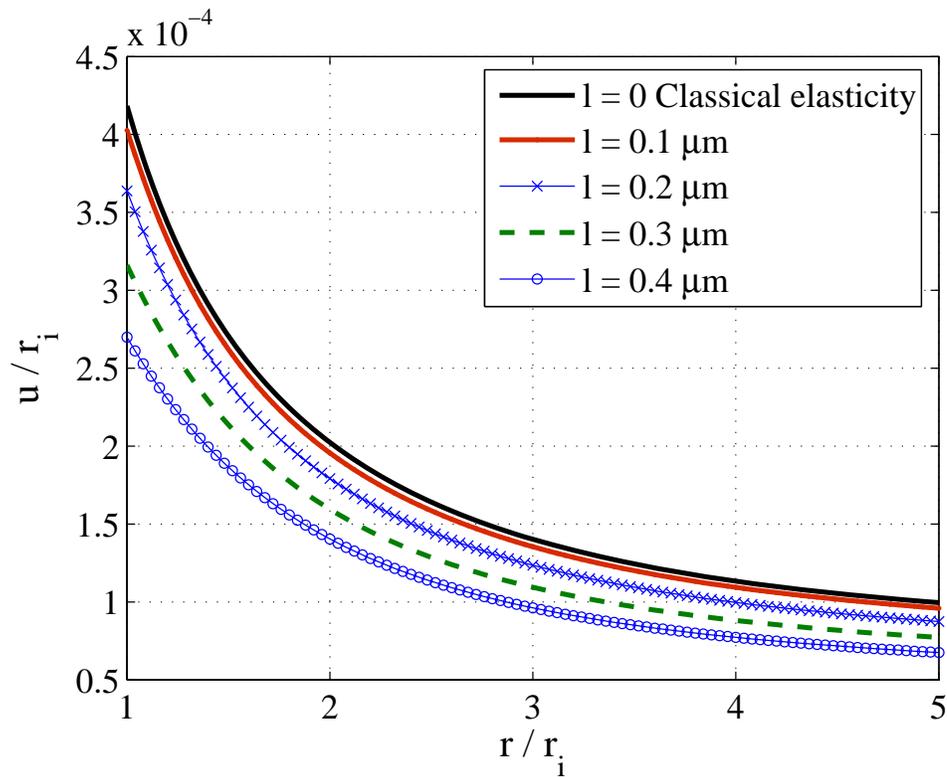

Figure 2: Non-dimensional radial displacement along a FG micro-cylinder radius for some values of $l$.

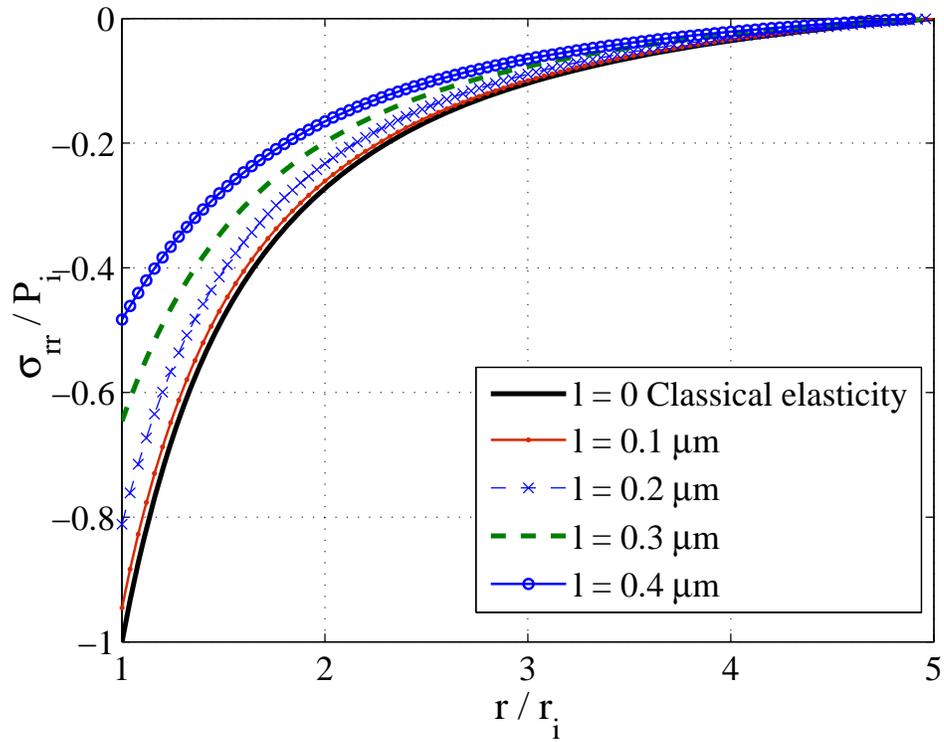

Figure 3: Non-dimensional radial stress distribution in a FG micro-cylinder along the radial direction for some values of material characteristic length parameter

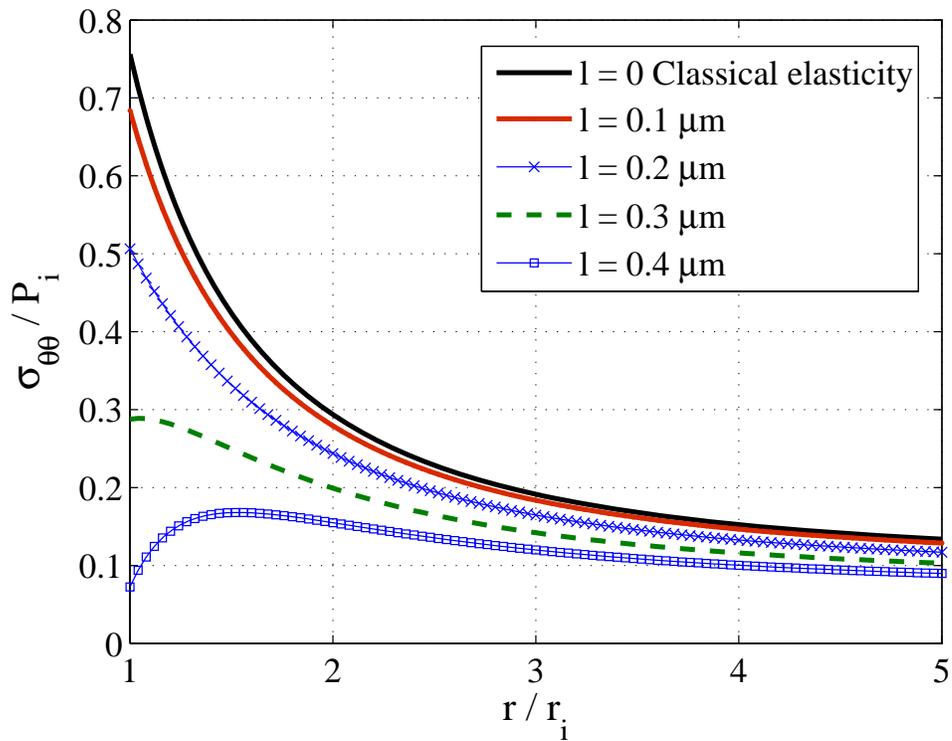

Figure 4: Non-dimensional tangential stress distribution in a FG micro-cylinder along the radial direction for some values of material characteristic length parameter.

Fig. 5- Fig. 7 show the strain gradient elasticity solution for non-dimensional displacement field as well as the radial and tangential stress distributions along the radial direction for some values of FG power index *n* in the FG micro-cylinder. In these figures the value of material characteristic length parameter is assumed to $l = 0.1 \ \mu m$. It can be understood from these figures that the value of *n* has a significant effect on the radial and tangential stress distributions in FG micro-cylinders. Furthermore, it can be seen that increasing FG power index *n* decreases the maximum radial and tangential stresses. For example, by increasing *n* from -1.5 to 1.5 non-dimensional maximum radial and tangential stresses decrease %8.4 and %88, respectively.

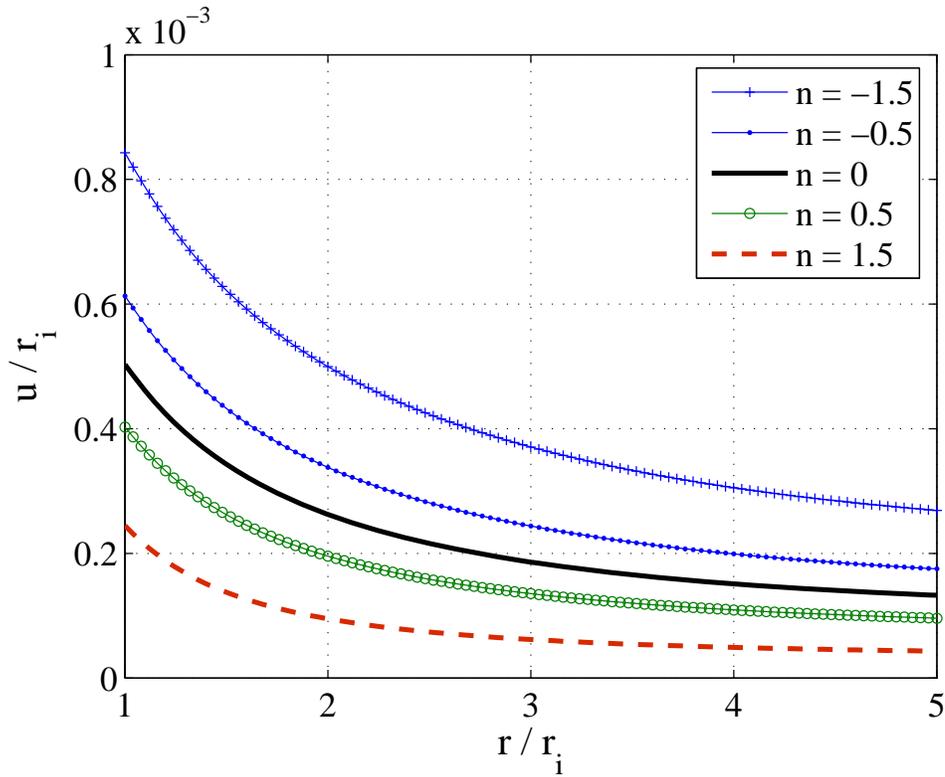

Figure 5: Non-dimensional radial displacement along the FG micro-cylinder radius for different values of FG power index *n*.

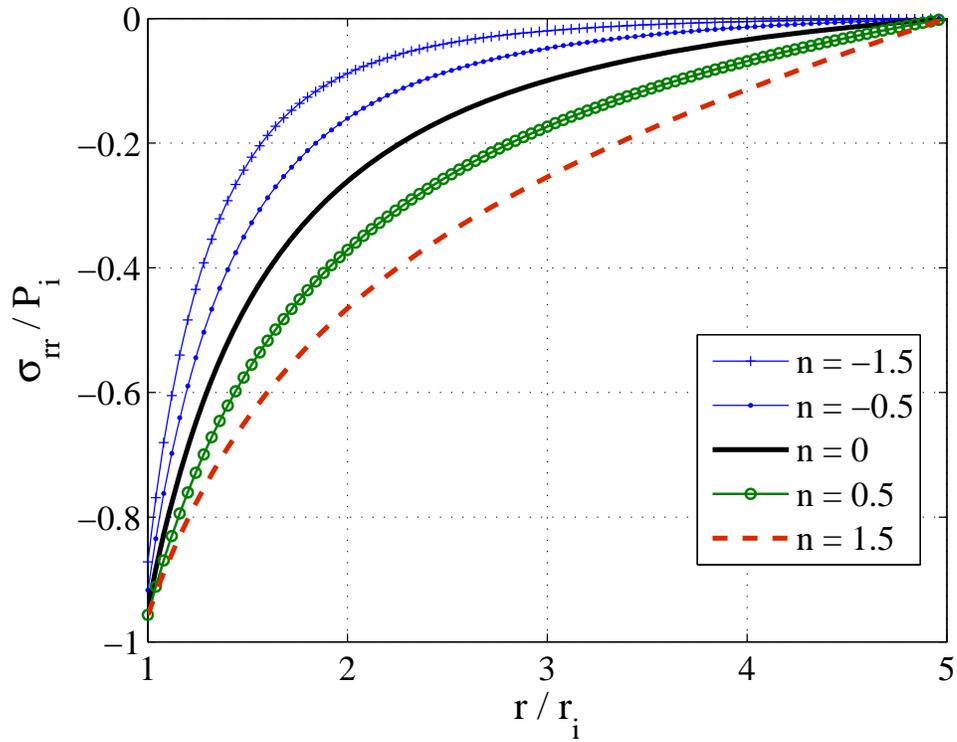

Figure 6: Non-dimensional radial stress distribution in a FG micro-cylinder along the radial direction for some values of FG power index *n*.

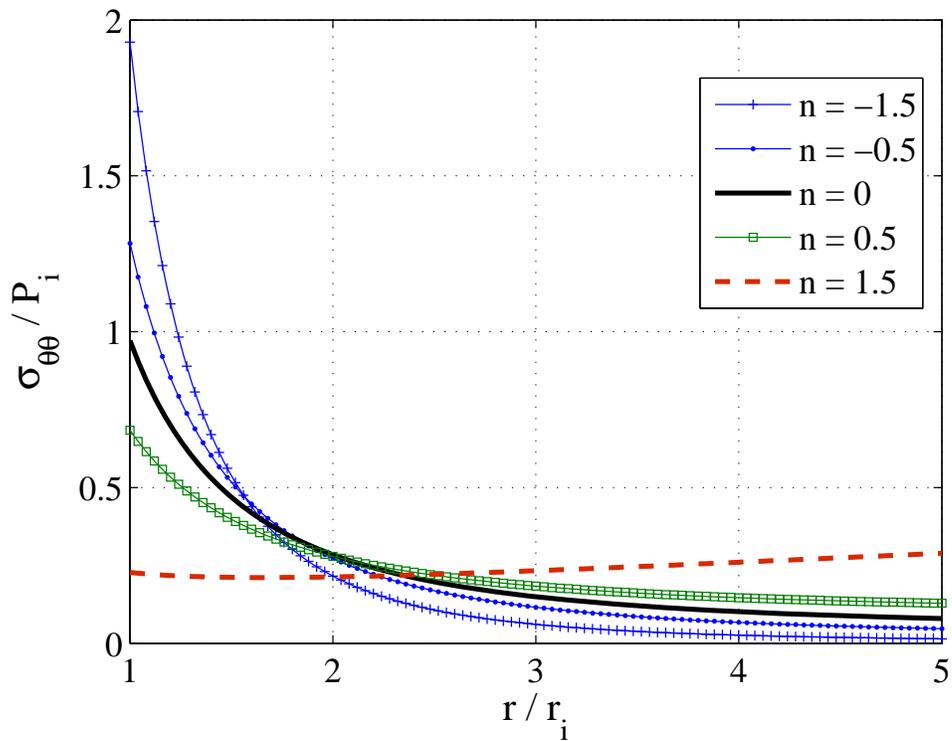

Figure 7: Non-dimensional tangential stress distribution in a FG micro-cylinder along the radial direction for some values of FG power index *n*.

## 5. Conclusion

Strain gradient elasticity formulation for axisymmetric FG micro-cylinders was presented. A power law distribution was assumed for variation of material properties in radial direction. A series form was presented for the solution of the governing equation. The effect of the characteristic length parameter as well as the effect of FG power index on the elastic deformation of FG micro-cylinder was studied. Results show that the characteristic length parameter has a significant effect on the stress distribution of FG micro-cylinders. Increase of material length parameter decreases the maximum radial and tangential stresses. Furthermore, it is shown that the FG power index $n$ has a significant effect on the radial and tangential stress distributions in FG micro-cylinders. It is shown that increasing FG power index $n$ decreases the maximum radial and tangential stresses.